%% file: main.tex
\documentclass[
  reprint,
  superscriptaddress,
  amsmath,amssymb,
  aps,
  prb,
]{revtex4-2}

\usepackage{graphicx}
\usepackage{dcolumn}
\usepackage{bm}
\usepackage{siunitx}
\usepackage{ragged2e}
\newcounter{subfigure}
 
\newcommand{\subfiganchor}[1]{\refstepcounter{subfigure}\label{#1}}
\newcommand{\figsub}[2]{Fig.~\hyperref[#2]{\ref*{#1}(\ref*{#2})}}
\usepackage{fancyhdr}
\usepackage{afterpage}      
\newlength{\PRLHeadsepFirst}
\newlength{\PRLHeadsepOther}
\usepackage[colorlinks=true,allcolors=blue]{hyperref}
\usepackage[section]{placeins}

\newcommand{\FigureSub}[2]{Figure~\hyperref[#2]{\ref*{#1}(\ref*{#2})}}

\begin{document}

\title{Coherent and dissipative coupling of multiple photon–magnon hybrids\\ in a chiral microwave resonator}

\author{Supanat Chantarattana}
\email{supanat.chantarattana.21@ucl.ac.uk}
\affiliation{Department of Physics and Astronomy, University College London, London WC1E 6BT, United Kingdom}
\affiliation{London Centre for Nanotechnology, University College London,  17-19 Gordon Street, London, WC1H 0AH, United Kingdom}
\author{Charlie W. F. Freeman}
\affiliation{London Centre for Nanotechnology, University College London,  17-19 Gordon Street, London, WC1H 0AH, United Kingdom}

\author{Daniel Prestwood}
\affiliation{London Centre for Nanotechnology, University College London,  17-19 Gordon Street, London, WC1H 0AH, United Kingdom}
\affiliation{Department of Electronic and Electrical Engineering, University College London, London WC1E 7JE, United Kingdom}

\author{Nimisha Arora}
\affiliation{London Centre for Nanotechnology, University College London,  17-19 Gordon Street, London, WC1H 0AH, United Kingdom}

\author{Vincent Castel}
\affiliation{IMT Atlantique, Technopole Brest-Iroise, CS 83818, 29238 Brest Cedex 3, France}

\author{Hidekazu Kurebayashi}
\email{h.kurebayashi@ucl.ac.uk}
\affiliation{London Centre for Nanotechnology, University College London,  17-19 Gordon Street, London, WC1H 0AH, United Kingdom}
\affiliation{Department of Electronic and Electrical Engineering, University College London, Roberts Building, London, WC1E 7JE, United Kingdom}
\affiliation{Center for Science and Innovation in Spintronics, Tohoku University, 2-1-1, Katahira, Sendai, 980-8577 Japan}
\affiliation{Institute for Materials Research, Tohoku University, 2-1-1, Katahira, Sendai, 980-8577 Japan}
\affiliation{WPI Advanced Institute for Materials Research, Tohoku University, 2-1-1, Katahira, Sendai 980-8577, Japan}

\begin{abstract}
We demonstrate control between level repulsion and level attraction of photon–magnon coupling spectra in a chiral microwave-photon resonator loaded with a \SI{2}{mm} YIG sphere. By systematically varying the coupling mode volume via the YIG position and the dc-field orientation, we tune the ratio between coherent and dissipative coupling strengths, $J/\Gamma$, and thereby switch between level repulsion and level attraction in the hybrid spectra. Full-wave electromagnetic and magnetization-dynamics simulations reveal multiple photon–magnon hybrid modes with distinct spatial overlap factors, which quantitatively account for gap sizes and their angular dependence. Strong nonreciprocity appears near $J/\Gamma\approx1$, yielding \SI{28.6}{dB} isolation ratio in agreement with a unified single-bath input--output model and full-wave simulations. Angular dependence of spectral characterization confirms a picture of the chirality-selection rule and our method of geometric tuning might be useful for considering reconfigurable on-chip isolators.
\end{abstract}

\setlength{\headsep}{\PRLHeadsepFirst}
\maketitle
\afterpage{\setlength{\headsep}{\PRLHeadsepOther}}

Cavity magnonics is the field that studies the interaction between magnons and confined photon modes in microwave or optical cavities, providing a versatile platform for exploring the use of future quantum information processing and magnonics applications \cite{LachanceQuirion2019APEX,Rameshti_PhysRep20221,Awschalom_IEEE2021}. When the coherent coupling rate between a collective magnon mode and a cavity photon mode exceeds their respective loss rates, the system enters a strong coupling regime, leading to level repulsion (avoided crossing) in their spectra \cite{Soykal_PRL2010,Huebl2013PRL}. Owing to a large spin density in magnetically ordered systems, which collectively enhances the coupling strength proportional to $\sqrt{N}$~\cite{Dicke_PR1954} ($N$ is the number of spins), the strong coupling in photon-magnon hybrids has been demonstrated in a number of different platforms, including cavity resonators coupled to ferrimagnetic spheres and films \cite{Huebl2013PRL,Bhoi_JAP2014,Tabuchi2014PRL,Zhang2014PRL,Flower2019NJP,Lambert2015JAP,Lee_PRL2023} and to noncollinear magnets hosting topological spin textures~\cite{Abdurakhimov_PRB2019,Khan_PRB2021}, on-chip resonators coupled to lithographically patterned ferromagnets~\cite{Hou_PRL2019,Li_PRL2019,Golovchanskiy_SciAdv2021,Golovchanskiy_PRAppli2021} and planar superconducting resonators~\cite{Morris_SciRep2017,Haygood_PRAppli2021,Ghirri_PRAppl2023,Zollitsch_NComm2023}. Furthermore, a magnon and a superconducting qubit are coherently coupled via a cavity mode \cite{Tabuchi_Science2015,Quirion_Science2020}, enabling the counting of individual magnon and photon numbers through discrete changes in the spectral features associated with transitions of the state population. 
  
If photon and magnon modes share a common radiative environment, a dissipative coupling channel  would emerge \cite{Harder2018PRL,Bhoi2019PRB,Grigoryan2019PRB,Rao2019NJP,Yang2019PRApp,Bourcin_PRApp2024,Wang_JAP2020} and, once it dominates over coherent coupling, leads to level attraction in the spectrum. 
Each of the coherent and dissipative coupling channels has its own oscillation phase, and they interfere strongly when their coupling strengths are comparable~\cite{Wang2019PRL}.
Crucially, this interference depends on the direction of microwave propagation. Nonreciprocity arising from this mechanism  has been demonstrated in cavity-magnonics devices based on on-chip cross-bar transmission lines~\cite{Wang2019PRL,Zhao_PRAppli2020,Qian_Ncomm2025} and chiral resonators~\cite{Zhang2020PRApp}. Nonreciprocal magnon polaritons have also been realized in other chiral-resonator geometries~\cite{Bourhill_PRAppli2023,Mita_PRAppli2025,Ardisson_arXiv2025}. Furthermore, nonreciprocal microwave transmission in a two-cavity magnonic system exploiting magnon Kerr nonlinearity has been theoretically proposed~\cite{Kong_PRAppli2019}. Compared to conventional ferrite-based devices~\cite{Pozar2011}, cavity-magnonics architectures offer potential advantages in terms of reconfigurability, on-chip integration, broadband operation, and large isolation ratios. This motivates systematic exploration of nonreciprocity across diverse photon–magnon platforms.

\begin{figure*}[!t]
  \centering
  \includegraphics[width=1\textwidth]{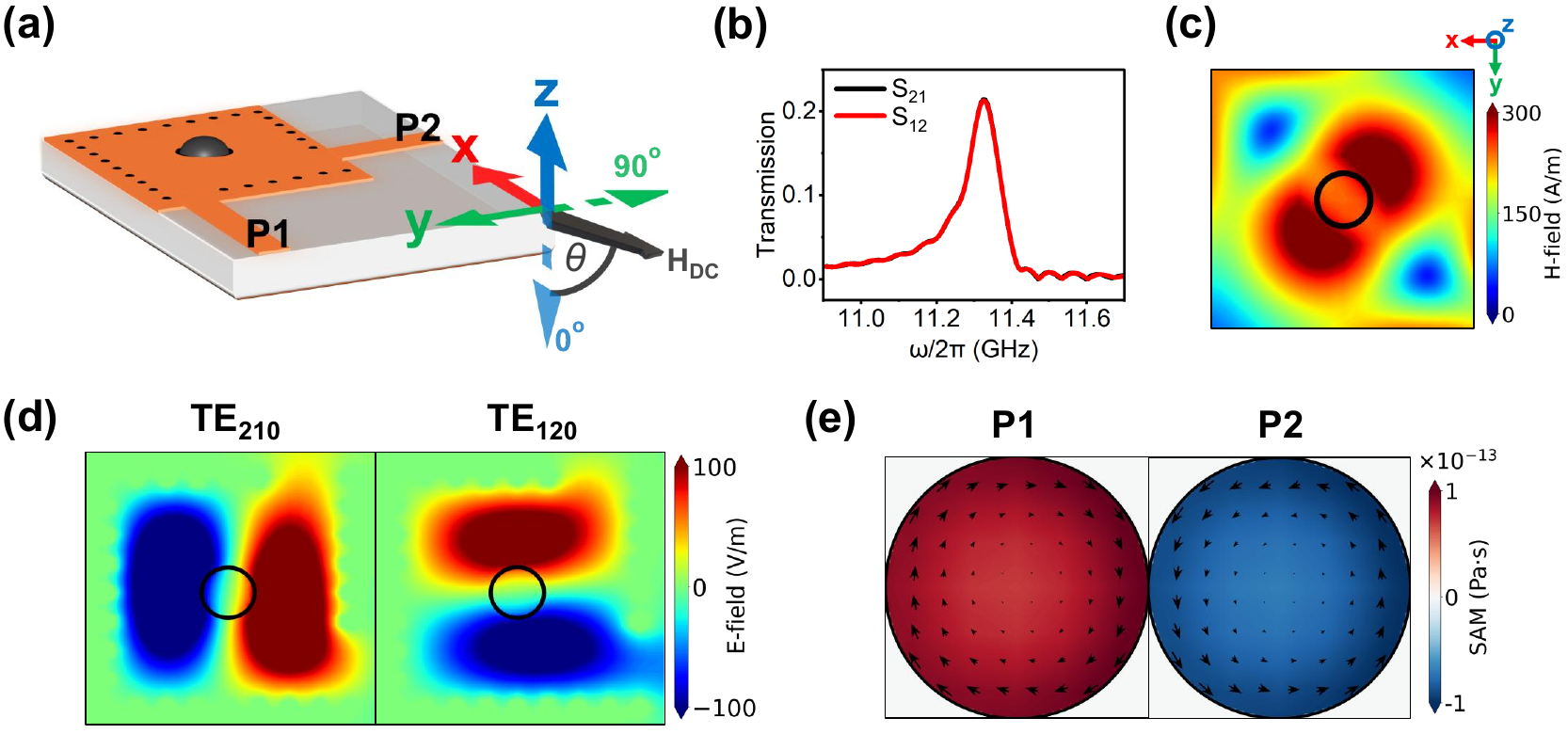}
  \setcounter{subfigure}{0}
  \subfiganchor{fig:1a}
  \subfiganchor{fig:1b}
  \subfiganchor{fig:1c}
  \subfiganchor{fig:1d}
  \subfiganchor{fig:1e}
\caption[Chiral SIW resonator and photon chirality.]%
  {\justifying
   (a) Substrate-integrated waveguide (SIW) resonator with a \SI{2}{mm} YIG sphere at the cavity center; the in-plane bias $\mathbf{H}_{\mathrm{DC}}$ lies between $-z$ and $-y$
with $\theta=0^\circ$ for $\mathbf{H}_{\mathrm{DC}}\parallel-z$
and $\theta=90^\circ$ for $\mathbf{H}_{\mathrm{DC}}\parallel-y$.
   (b) Simulated transmission spectra $S_{21}$ and $S_{12}$.
   (c) Microwave magnetic-field magnitude $\lvert \mathbf{H} \rvert$ on the top $xy$ plane at resonance; the field is strongly concentrated around the sphere position (black circle).
   (d) Electric-field patterns of the orthogonal TE$_{210}$ and TE$_{120}$ modes; a $\pm \pi/2$ superposition realizes clockwise/counter-clockwise photon states.
  (e) Spin-angular-momentum (SAM) maps for ports P1 and P2 showing opposite handedness (color: SAM density; arrows: Poynting vector).}

  \label{Fig:1}
\end{figure*}

In this work, we use on-chip planar microwave cavity resonator with circular polarization of photon modes, as demonstrated previously by Zhang et al.~\cite{Zhang2020PRApp}. Unlike the previous report~\cite{Zhang2020PRApp}, we observed multiple avoided crossings, suggesting the existence of additional magnon modes coupled to cavity modes. Our numerical simulations that solved coupled electromagnetic and magnetization dynamics revealed the spatial profile of the two magnon modes interacting with the cavity photons, thereby explaining the difference in coupling strength. The mode volume overlap clarified by the simulations accounts for the out-of-plane angular dependence of the gap size for both hybridized modes. Furthermore, going beyond the regime explored in the previous study~\cite{Zhang2020PRApp}, we observe the strength of nonreciprocity is suppressed by increasing the coherent coupling strength. We discuss this by using the relative strength of dissipative and coherent couplings ($J$ and $\Gamma$ respectively) that was controlled by systematic changes of the sphere position, providing the nonreciprocity of microwave transmission in terms of $J/\Gamma$. We experimentally show that the nonreciprocity is maximized near $J/\Gamma \approx 1$ and disappears in the strong coupling regime where $J/\Gamma \gg 1$.

We first describe our CST simulations to explain resonance modes in an on-chip cavity (see more details in Supplementary Material). Calculated transmission spectra $S_{21}$ and $S_{12}$ [\figsub{Fig:1}{fig:1b}] perfectly overlap each other and show a resonance mode at \SI{11.33}{GHz}. The magnitude of microwave magnetic field $\lvert \mathbf{H} \rvert$ on the top $xy$ plane is concentrated around the center of the cavity where we placed our YIG sphere, as indicated by a circle in \figsub{Fig:1}{fig:1c}. Chirality in our photon modes arises because two nearly degenerate linearly-polarized modes, $\mathrm{TE}_{210}$ and $\mathrm{TE}_{120}$ shown in \figsub{Fig:1}{fig:1d}, are excited with a $\pm\pi/2$ relative phase due to the two microwave ports orthogonally connected to the resonator~\cite{Zhang2020PRApp}. Their superposition yields photon modes with opposite circular polarization, controllable by the microwave propagation direction from either Port 1 (P1) or Port 2 (P2).

To quantify photon chirality and local energy flow in the resonator, we evaluate the Poynting vector $\boldsymbol{\mathcal P}$ and the spin angular-momentum (SAM) density of the electromagnetic field $\mathbf{s}$: $\boldsymbol{\mathcal P} = (1/2)\operatorname{Re}\!\left(\mathbf{E}^* \times \mathbf{H}\right)$, $\mathbf{s} = (1/4\omega)\operatorname{Im}\!\left[\epsilon_0\,\mathbf{E}^* \times \mathbf{E}
+ \mu_0\,\mathbf{H}^* \times \mathbf{H}\right],$ where $\mathbf{E}$ and $\mathbf{H}$ are the complex electric and magnetic fields \cite{Jackson1999,Bliokh2017PRL}. 
At microwave frequencies, $\mathbf{s}$ is dominated by the magnetic contribution, $\mathbf{s} \propto \operatorname{Im}(\mathbf{H}^* \times \mathbf{H})$. In \figsub{Fig:1}{fig:1e}, we visualize $\boldsymbol{\mathcal P}$ as arrows and $\mathbf{s}$ as color in the YIG sphere area, to demonstrate the sign change in $\mathbf{s}$ controlled by the reversal of microwave propagation. This is one of our key experimental controls in the present study.

\begin{figure*}[!t]
  \centering
  \includegraphics[width=1\textwidth]{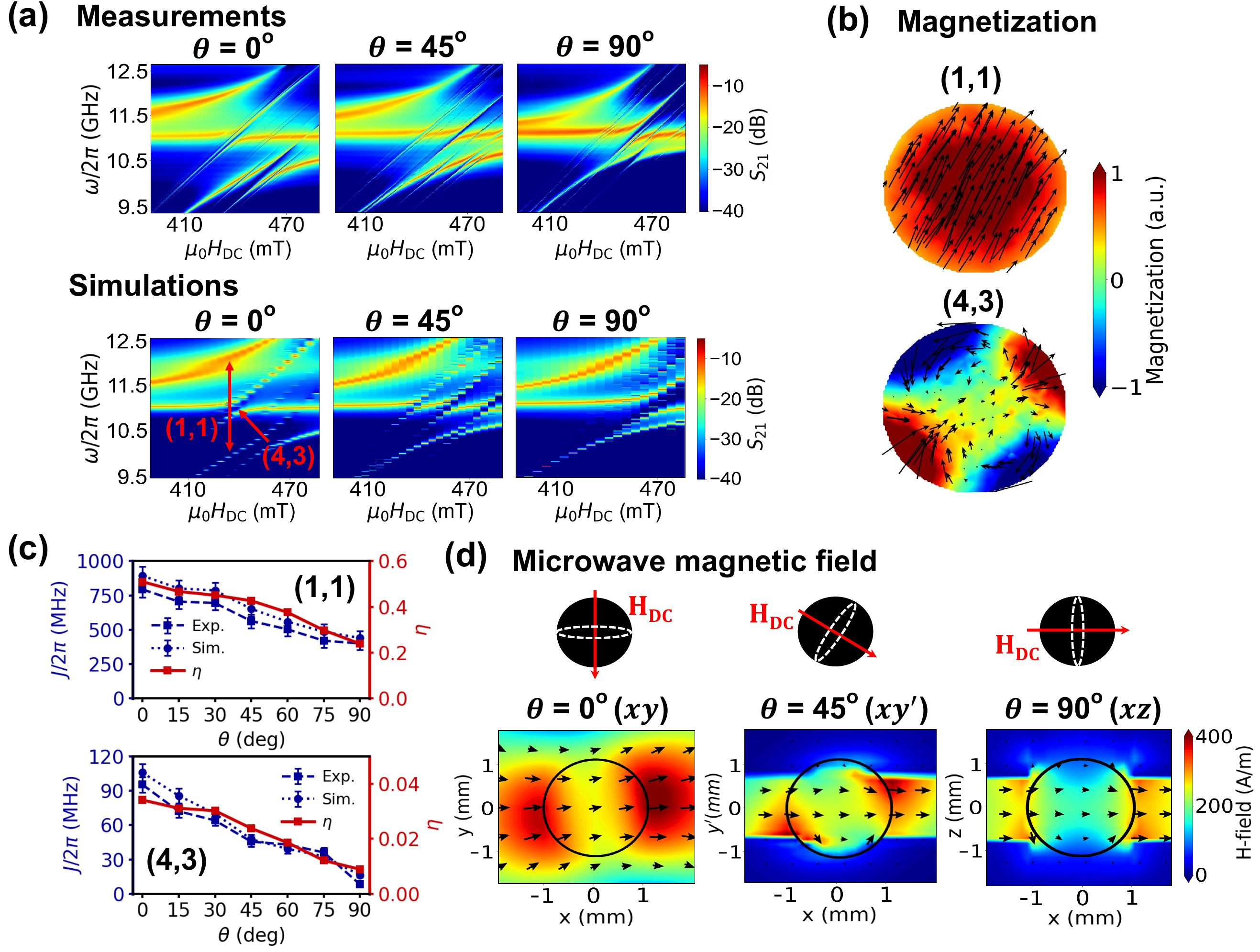}
  \setcounter{subfigure}{0}
  \subfiganchor{fig:2a}
  \subfiganchor{fig:2b}
  \subfiganchor{fig:2c}
  \subfiganchor{fig:2d}
\caption[Angle dependence of coupling, mode textures, and microwave fields.]%
{\justifying
(a) Measured (top) and COMSOL-simulated (bottom) transmission versus frequency and bias angle $\theta$, showing two anticrossings corresponding to the $(1,1)$ and $(4,3)$ magnon modes.
(b) COMSOL maps of the dynamic magnetization for the $(1,1)$ and $(4,3)$ modes in the central cross-section of the \SI{2}{mm}-diameter YIG sphere.
(c) Angular dependence of the coherent coupling $J/2\pi$ (left axis) and the overlap coefficient $\eta$ (right axis) versus $\theta$.
(d) CST maps of the microwave magnetic field at $\theta = 0^\circ, 45^\circ$, and $90^\circ$. To highlight the transverse driving components, the fields are plotted in planes perpendicular to the applied dc magnetic field $\mathbf{H}_{\mathrm{DC}}$, as indicated visually by the 3D sphere insets. The color encodes $|\mathbf H|$ in A/m; arrows show in-plane field direction and relative magnitude. The black circle marks the sphere boundary. For $\theta=0^\circ$ the section is the $xy$ plane; for $\theta=90^\circ$ it is the $xz$ plane; for $\theta=45^\circ$ it is the rotated plane $y'=(y+z)/\sqrt{2}$, giving a mixed projection of the cavity field.
}
  \label{Fig:2}
\end{figure*}

\FigureSub{Fig:2}{fig:2a} summarizes measured and simulated transmission spectra for the polar angle $\theta = 0^\circ$, $45^\circ$, and $90^\circ$ (defined in Fig.1a) in the applied magnetic field ($\mathbf{H}_{\mathrm{DC}}$) near mode crossing. For $\theta = 0^\circ$, multiple avoided crossings can be recognized, from which we focus on one with the largest gap (we labeled as (1,1)) as well as another with a much smaller gap, denoted as (4,3) in one of the plots in \figsub{Fig:2}{fig:2a}. Labels $(n,m)$ specify the spinwave spatial profile with the two integers representing the nodal structure of the dynamic magnetization in the polar and azimuthal directions (see Supplemental Material), following the spherical magnetostatic mode framework established by Walker, Fletcher, and Bell \cite{Walker1957, FletcherBell1959}. The spatial profile of the two specific spinwave modes is plotted in \figsub{Fig:2}{fig:2b} which we used for quantitative analysis later. Our LLG-incorporated COMSOL simulations with appropriate device and magnetic parameters help us identify these specific modes that couple with the cavity resonator mode. The simulation results display excellent agreement with experiments as shown in \figsub{Fig:2}{fig:2a} for corresponding $\theta$ values. 
Both experimental and simulation spectra retain the large and small hybridization gaps for $0^\circ \leq \theta \leq 90^\circ$ but both gap sizes shrink as $\theta$ is increased. 

For quantitative analysis, we fit both the experimental and
COMSOL-simulated transmission spectra using a unified
single-cavity--single-magnon input--output model. The coherent and
dissipative cavity--magnon interactions are treated within a common
radiative bath~\cite{Metelmann_PRX2015,Wang2019PRL}, such that the dissipative coupling is constrained by
$\Gamma=R\sqrt{\kappa_c\kappa_m}$, where $\kappa_c$ and $\kappa_m$
are the external coupling rates of the cavity and magnon modes,
respectively, and $R$ is a real direction-independent constant.
The resulting transmission expression is derived in the Supplemental
Material. With the cavity and magnon detunings defined as
$\Delta_c=\omega-\omega_c$ and $\Delta_m=\omega-\omega_m$,
respectively, and defining
$X_c=\Delta_c+i\gamma_c$, $X_m=\Delta_m+i\gamma_m$, and
$W=J-i\Gamma e^{i\Phi}$, the forward and backward transmission
coefficients are given by:

\begin{widetext}
\begin{equation}
S_{21(12)} =
\frac{
W^2-X_cX_m
-\sqrt{\kappa_c\kappa_m}\left(1+e^{i\Phi}\right)W
+\kappa_c\kappa_m\left[1-i\left(1+e^{i\Phi}\right)\right]
-\kappa_c(1+i)X_m
-\kappa_m\left(e^{i\Phi}+i\right)X_c
}
{
(\kappa_c+\gamma_c-i\Delta_c)
(\kappa_m+\gamma_m-i\Delta_m)
+\left(J-i\Gamma e^{i\Phi}\right)^2
},
\label{eq:s21_gen}
\end{equation}
\end{widetext}

where $\kappa_{c,m}$ denote the external coupling rates,
$\gamma_{c,m}$ the intrinsic loss rates of the cavity ($c$) and
magnon ($m$), and $\omega_{c,m}$ their corresponding resonance
frequencies. The dissipative coupling is determined by the common-bath
relation introduced above.
The direction-dependent phase is represented by $\Phi=0$ for forward
($S_{21}$) propagation and $\Phi=\pi$ for backward ($S_{12}$)
propagation.
\FigureSub{Fig:2}{fig:2c} includes extracted $J/2\pi$ from experiments and numerical calculations plotted in \figsub{Fig:2}{fig:2a} for the two hybridized states, where all $J/2\pi$ plots represent a decrease with increasing $\theta$. Phenomenologically, the coherent coupling strength can be estimated from the cavity--magnon coupling model of Ref.~\cite{Flower2019NJP}:\begin{equation}
  J = \frac{\gamma }{2}\eta
    \sqrt{\frac{\mu\mu_0  n_s\hbar \omega_c}{g\mu_B}},
  \label{eq:J_eta_relation}
\end{equation}
where $\gamma = g\mu_B/\hbar$ is the gyromagnetic ratio, $g$ is the Land\'e $g$-factor, $\mu_B$ is the Bohr magneton, $\mu$ is the magnetic moment per spin in the medium, $\mu_0$ is the vacuum permeability, and $n_s$ is the spin density. The dimensionless overlap factor $\eta\in[0,1]$ can be calculated from the spatial overlap between the unhybridized cavity microwave magnetic field and magnon mode as~\cite{Lambert2015JAP}
\begin{equation}
  \eta = \left|
  \int_{\mathrm{sphere}}
  \frac{\mathbf{H}(\mathbf{r}) \!\cdot\! \mathbf{M}(\mathbf{r})}
       {H_{\max}\,M_{\max}\,V_{\mathrm{YIG}}}\,\mathrm{d}V
  \right|,
  \label{eq:eta_def}
\end{equation}
where $\mathbf{H}(\mathbf r)$
is the cavity microwave magnetic field as a function of position $\mathbf{r}$, $\mathbf{M}(\mathbf{r})$ is the dynamic magnetization of the magnon mode, $H_{\max}$ and $M_{\max}$ are their respective maxima within the YIG sphere, and $V_{\mathrm{YIG}}$ is the sphere volume. $\mathbf{H}(\mathbf r)$ of the uncoupled cavity mode profile is obtained from full-wave simulations, while the bare dynamic magnetization $\mathbf{M}(\mathbf{r})$ is calculated analytically using the exact magnetostatic modes for a ferrite sphere \cite{FletcherBell1959} (see Supplemental Material for details on the Walker-mode labeling). \FigureSub{Fig:2}{fig:2b} represents a constant sign of $\mathbf{M}$ across the sphere for the $(1,1)$ mode, explaining large $\eta$. In contrast, the higher-order $(4,3)$ mode contains internal nodes that reduce $\eta$, and hence $J$. A good scaling between $\eta$ and $J$ shown in \figsub{Fig:2}{fig:2c} further suggests that $\eta$ is the primary cause of their $\theta$ dependence. We note that $\Gamma$ remains considerably smaller than $J$ across the measured angular range, as summarized in the Supplemental Material.

\begin{figure*}[!t]
  \centering
  \includegraphics[width=1\textwidth]{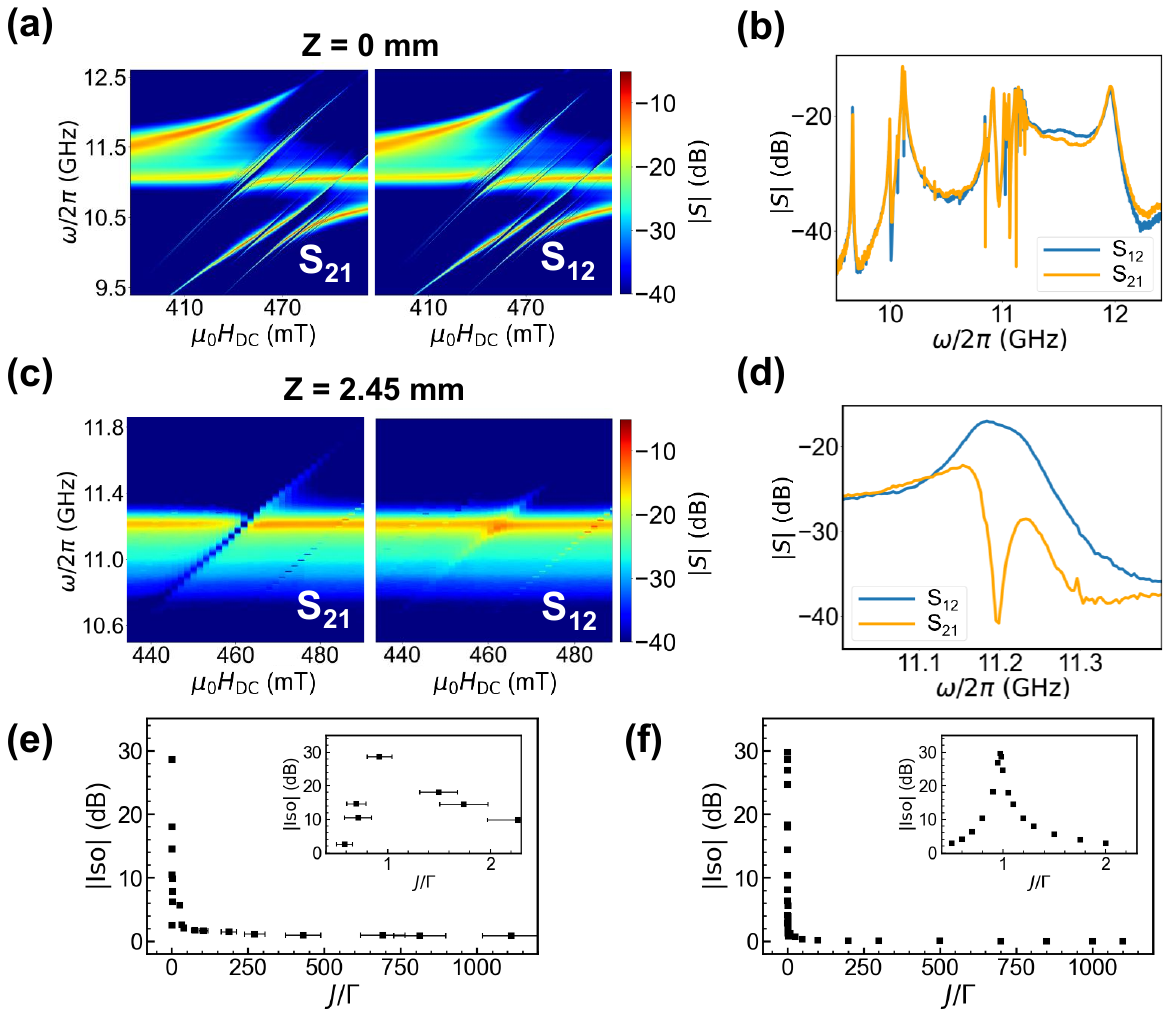}
  \setcounter{subfigure}{0}
  \subfiganchor{fig:3a}
  \subfiganchor{fig:3b}
  \subfiganchor{fig:3c}
  \subfiganchor{fig:3d}
  \subfiganchor{fig:3e}
  \subfiganchor{fig:3f}
\caption[Height-controlled nonreciprocity linked to $J/\Gamma$.]%
{\justifying
Geometric control of nonreciprocity by varying the sphere height $Z$.
(a,c) Measured $S_{21}$ and $S_{12}$ maps at $Z=0$\,mm and $Z=2.45$\,mm.
(b,d) Frequency cuts taken at $\mu_0H_{\mathrm{DC}} = 444.3~\mathrm{mT}$ and $\mu_0H_{\mathrm{DC}} = 460.5~\mathrm{mT}$, respectively, showing the forward ($S_{21}$) and backward ($S_{12}$) transmission spectra.
(e,f) Isolation ratio $\mathrm{Iso}$ versus $J/\Gamma$ from experiment and unified-model calculation (insets: small-$J/\Gamma$ zoom).}
  \label{Fig:3}
\end{figure*}

So far, we have presented the results and analysis for the YIG sphere positioned within the microwave resonator, corresponding to a vertical displacement of $Z = 0$. We next demonstrate a transition from reciprocal to nonreciprocal transmission, realized by controlling $J/\Gamma$ with $Z$. From $Z = 0$ where $J \gg \Gamma$, increasing $Z$ causes a change of electromagnetic couplings, as the YIG sphere exits the localized antinode of the cavity's modes, which decreases $\eta$, hence $J$ reduction. The $Z$ dependence of $\Gamma$ differs from that of $J$ because, in the unified model, $\Gamma$ is determined by the external cavity and magnon coupling rates, whereas $J$ is governed primarily by the cavity--magnon mode overlap. Extended data mapping the continuous $Z$ dependence of these parameters are provided in the Supplemental Material.

Figure~\ref{Fig:3} summarizes the nonreciprocity observed in our photon-magnon hybrids. Figures ~\hyperref[fig:3a]{\ref*{Fig:3}(\ref*{fig:3a})} and \hyperref[fig:3b]{\ref*{Fig:3}(\ref*{fig:3b})} represent the spectral response at $Z=0$ for two microwave propagation directions ($S_{21}$ and $S_{12}$), suggesting negligible nonreciprocity with the coherent coupling feature, supported by $J/\Gamma = 1110 \pm 90$ extracted using Eq.~\eqref{eq:s21_gen}. For $Z=2.45$\,mm shown in Figs.~\hyperref[fig:3c]{\ref*{Fig:3}(\ref*{fig:3c})} and \hyperref[fig:3d]{\ref*{Fig:3}(\ref*{fig:3d})}, the level-attraction structure appears in the $S_{21}$ spectra, accompanied by nonreciprocal transmission. In this case, the system is tuned close to the balance condition between coherent and dissipative coupling, with $J/\Gamma = 0.92 \pm 0.12$.

Observed nonreciprocity can be explained by the term of $-2iJ\,\Gamma e^{i\Phi}$ that appears when we expand the parenthesis in the denominator in Eq.~\eqref{eq:s21_gen}. This term therefore represents the interference of the coherent and dissipative couplings and is proportional to each coupling strength. Another key ingredient of this nonreciprocal mechanism is the
direction-dependent sign of the dissipative coupling term, represented by $\Phi=0$ and $\Phi=\pi$ for the two propagation directions. This is expected because reversing the microwave propagation direction reverses the driving current, and by Ampère’s law the generated magnetic field is an odd function of the current, producing the $\pi$ phase change. As described below, our observation matches well with this simple model.

For quantitative discussions, we define the isolation ratio as $|\mathrm{Iso}| = |S_{21}-S_{12}|$ (in dB) and plot its experimental values as a function of $J/\Gamma$ in Fig.~\hyperref[fig:3e]{3(e)}. We observe the largest $\mathrm{Iso}$ when $J/\Gamma$ is close to unity and a monotonic decrease is present as $J/\Gamma$ increases beyond 1. This suppression of nonreciprocity deep in the strong coupling regime, where $J/2\pi$ approaches 800 MHz, has not been systematically explored in previous work; for comparison, the chiral resonator geometry of Zhang et al. \cite{Zhang2020PRApp} operated up to $J/2\pi \approx 100$ MHz and did not investigate the large-$J/\Gamma$ regime. The highest amplitude of $\mathrm{Iso}$ reaches 28.6 dB at $J/\Gamma = 0.92 \pm 0.12$, which we here compare with earlier reports. 
Among cavity-magnonic systems, Wang et al. demonstrated nonreciprocal microwave transmission with isolation ratios theoretically diverging at the zero-damping condition using a cross-bar cavity geometry \cite{Wang2019PRL}. Qian et al. later achieved up to 80 dB by introducing a tunable loop antenna to precisely control the magnon radiation damping~\cite{Qian_APL2020} and Wang et al. reported 46.2 dB in a system combining a split-ring resonator with an active gain circuit \cite{Wang_APL2023}. In a chiral resonator loaded with multiple YIG spheres, isolation of up to 46~dB has recently been reported~\cite{Ardisson_arXiv2025}. In non-magnonic microwave platforms, optomechanical implementations, including on-chip electromechanical circulators and reservoir-engineered circuits, have demonstrated isolation in the range of 10–15 dB \cite{Bernier_NComm2017,Peterson_PRX2017}. 

\begin{figure}[!t]
  \centering
  \includegraphics[width=1\linewidth]{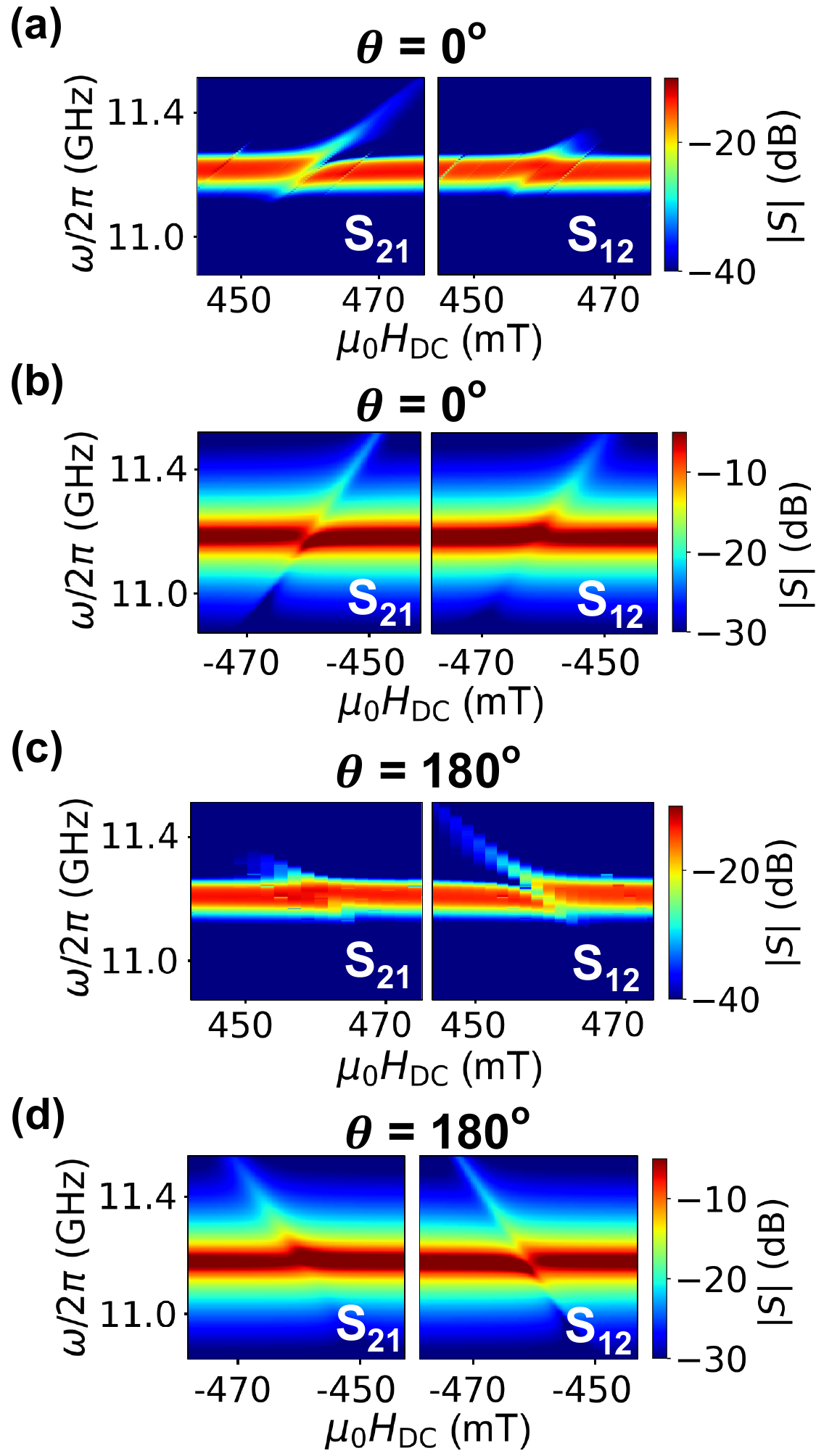}
  \setcounter{subfigure}{0}
  \subfiganchor{fig:4a}
  \subfiganchor{fig:4b}
  \subfiganchor{fig:4c}
  \subfiganchor{fig:4d}
  \caption[Chirality-selective coupling under field reversal.]
  {\justifying
  Chirality-selective coupling verified by reversing the applied dc bias field at a fixed sphere height of $Z=2.45$\,mm.
  (a,c) Experimental transmission maps measured for $\theta=0^\circ$ ($\mathbf H_{\mathrm{DC}}\parallel-\hat{\mathbf z}$) and $\theta=180^\circ$ ($\mathbf H_{\mathrm{DC}}\parallel+\hat{\mathbf z}$), respectively. (b,d) Corresponding transmission maps calculated using the unified single-bath input--output model of Eq.~\eqref{eq:s21_gen}.
  }
  \label{Fig:4}
\end{figure}

Our experimental results agree well with calculations using
Eq.~\eqref{eq:s21_gen}, which predict a maximum isolation of
approximately 29.7~dB near $J/\Gamma=0.97$, as shown in
Fig.~\hyperref[fig:3f]{3(f)}. The calculation uses
$\kappa_c/2\pi=\SI{4.1}{MHz}$,
$\gamma_c/2\pi=\SI{8.0}{MHz}$,
$\gamma_m/2\pi=\SI{0.196}{MHz}$, and
$J/2\pi=\SI{48}{MHz}$. In the unified model, $\Gamma$ is not varied
independently but is determined through
the common-bath relation
above; the ratio $J/\Gamma$ is therefore
controlled through the corresponding external coupling rates.
Both experiment and the unified-model calculation confirm that the
isolation is maximized near $J/\Gamma \approx 1$ and decreases as the
system moves away from the balanced regime, reconfirming that the
balance between coherent and dissipative coupling is the essential
requirement for maximum nonreciprocity. The small difference between
the experimental peak of 28.6~dB and the calculated value of
approximately 29.7~dB can be attributed to the fact that varying $Z$
simultaneously modifies several system parameters, including the
external cavity and magnon coupling rates and the cavity--magnon mode
overlap. Extended data mapping the continuous $Z$ dependence of
$\Gamma$ and the $J/\Gamma$ ratio are provided in the Supplemental
Material.

We next verify the chirality-selection rule governing the photon--magnon coupling by independently reversing the magnon chirality via inversion of the dc bias magnetic field ($\theta = 0^\circ$ to $180^\circ$) and the photon chirality via reversal of the microwave propagation direction.
As shown in the experimental transmission maps in \figsub{Fig:4}{fig:4a} and \figsub{Fig:4}{fig:4c}, a pronounced hybridization feature appears in $S_{21}$ for $\theta=0^\circ$ and in $S_{12}$ for $\theta=180^\circ$. The corresponding transmission maps calculated using the unified single-bath input--output model [Eq.~\eqref{eq:s21_gen}] are shown directly below in \figsub{Fig:4}{fig:4b} and \figsub{Fig:4}{fig:4d}, reproducing the experimentally observed chirality-selective coupling. This coupling is strongly suppressed when either the magnon or photon chirality alone is reversed, but is restored when both are flipped simultaneously. This behaviour is consistent with matched and mismatched handedness corresponding to $\Phi=0$ and $\Phi=\pi$, respectively, in Eq.~\eqref{eq:s21_gen}, confirming that the chirality-selection rule strictly governs the directional phase in our on-chip open cavity~\cite{Zhang2020PRApp,Wang2019PRL,Bourhill_PRAppli2023}.

In conclusion, we have demonstrated geometric control of coherent and dissipative photon--magnon coupling in an on-chip chiral microwave resonator loaded with a YIG sphere, providing a systematic route to tune the ratio $J/\Gamma$ and thereby switch between level repulsion and level attraction in the hybrid spectra. 
Full-wave electromagnetic and magnetization-dynamics simulations reveal the spatial profiles of multiple magnon modes coupled to the cavity photon mode, with the dimensionless overlap factor  quantitatively accounting for the difference in coherent coupling strengths and their angular dependence. 
By displacing the YIG sphere vertically from the cavity center, we continuously tuned $J/\Gamma$ from the strongly coherent regime ($J/\Gamma \gg 1$) to the balanced condition ($J/\Gamma \approx 1$), at which the interference between coherent and dissipative coupling channels is maximized and nonreciprocity peaks at an isolation ratio of 28.6~dB, in quantitative agreement with our unified single-bath input--output model and full-wave simulations.
The chirality-selection rule is examined and confirmed by independently changing the magnon and photon chirality in our experiments. 
This study presents details of experiments and analysis of cavity-magnonic platforms with controllable chirality and nonreciprocity, which might be useful for on-chip microwave transmission. 

\vspace*{6pt}

\textit{Acknowledgments} --
We acknowledge support by EPSRC under EP/X015661/1. D.P. is supported by the EPSRC and SFI Centre for Doctoral Training in Advanced Characterisation of Materials Grant Ref: EP/S023259/1. H.K. thanks the Leverhulme Trust for financial support via their research fellowship (RF-2024-317). We acknowledge IMT Atlantique for providing computational resources for the multiphysics simulations.

\vspace*{6pt}
\textit{Data availability} --
The data that support the findings of
this article are not publicly available. The data are available
from the authors upon reasonable request.

\bibliographystyle{apsrev4-2}   
\bibliography{ref}

\clearpage
\onecolumngrid

\setcounter{secnumdepth}{3}
\makeatletter
\renewcommand{\thesection}{\Roman{section}.}
\def\@seccntformat#1{\csname the#1\endcsname\quad}
\makeatother

\setcounter{equation}{0}
\setcounter{figure}{0}
\setcounter{table}{0}
\renewcommand{\theequation}{S\arabic{equation}}
\renewcommand{\thefigure}{S\arabic{figure}}
\renewcommand{\thetable}{S\arabic{table}}
\def\theHequation{S\arabic{equation}}
\def\theHfigure{S\arabic{figure}}
\def\theHtable{S\arabic{table}}

\newcommand{\MTfig}[1]{Main Text Fig.~#1}
\newcommand{\MTeq}[1]{Main Text Eq.~(#1)}
\makeatletter
\newcommand{\smcite}[1]{{\def\@auxout{-1}\cite{#1}}}
\makeatother

\begin{center}
  {\large\bfseries Supplementary Material for\\[2pt]
  Coherent and dissipative coupling of multiple photon--magnon hybrids\\[2pt]
  in a chiral microwave resonator\par}
  \vspace{8pt}
  Supanat Chantarattana, Charlie W. F. Freeman, Daniel Prestwood, Nimisha Arora,\\
  Vincent Castel, and Hidekazu Kurebayashi
\end{center}
\vspace{1.0em}

\section{Experimental geometry}
\input{sections/01-geometry/section.tex}

\FloatBarrier

\section{CST-simulated microwave magnetic field}

\input{sections/02-cst-fieldmaps/section.tex}
\FloatBarrier

\section{Identification of modes that couple to the cavity photon mode}

\input{sections/03-mode-id/section.tex}
\FloatBarrier

\section{Unified single-bath input--output model: derivation of \texorpdfstring{$S_{21}$}{S21} and \texorpdfstring{$S_{12}$}{S12}}

\input{sections/04-gen-io/section.tex}
\FloatBarrier

\section{Angular dependence of the dissipative coupling rate}
\input{sections/05-angle-dep/section.tex}
\FloatBarrier

\section{Height dependence of the coupling parameters}
\input{sections/06-height-dep/section.tex}
\FloatBarrier

\section*{References}
{\makeatletter\let\label\@gobble
\input{sm_refs.tex}}

\end{document}

%% file: sections/01-geometry/section.tex
\begin{figure}[!htb]
  \centering
  \includegraphics[width=0.85\linewidth]{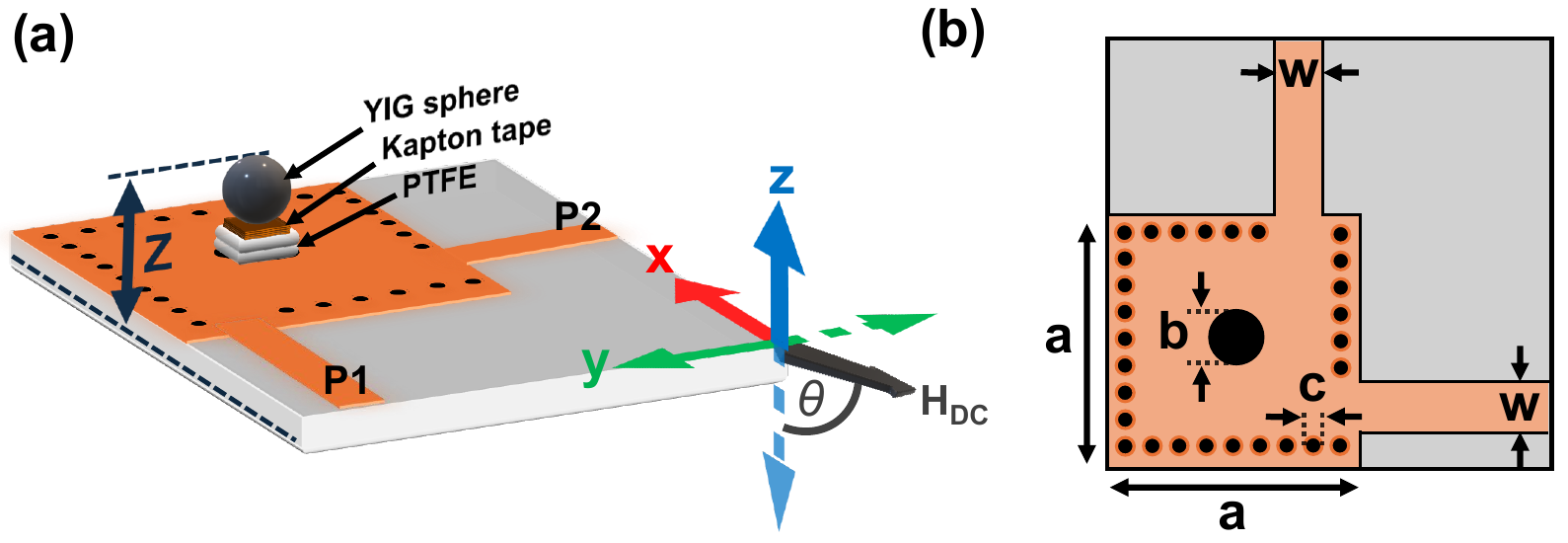}
  \setcounter{subfigure}{0}
  \subfiganchor{fig:S1a}
  \subfiganchor{fig:S1b}
\caption[Experimental geometry of the SIW–YIG platform.]
{\justifying
\textbf{Experimental geometry.}
(a) Schematic of the height-controlled SIW–YIG experiment. The chiral SIW resonator is measured in transmission between ports 1 and 2 (labeled P1 and P2). The external dc magnetic field $\mathbf H_{\mathrm{DC}}$ lies in the $yz$ plane
at angle $\theta$ to the $z$ axis. The YIG sphere is mounted on a spacer assembly comprising a \SI{1}{mm} PTFE block and \SI{0.05}{mm} Kapton layers, allowing its height $Z$ to be adjusted from \SI{0} to \SI{3}{mm} above the cavity surface.
(b) Top-view geometry of the cavity, showing key dimensions: $a=10.8$~mm, $b=2$~mm, $c=0.7$~mm, $w=1.75$~mm. The via holes, spaced \SI{1.15}{mm} apart, confine electromagnetic fields within the substrate.}
  \label{fig:S1}
\end{figure}

\noindent
The experimental platform is a substrate-integrated waveguide (SIW) cavity incorporating a \SI{2}{mm}-diameter YIG sphere positioned at the center of the cavity in the $xy$ plane \smcite{SM-Zhang2020PRApp}, as shown in \figsub{fig:S1}{fig:S1a}. The complex transmission spectra $S_{21}$ and $S_{12}$ were measured using a vector network analyser (Rohde \& Schwarz ZNB40) in a two-port configuration. The VNA was connected to the resonator ports using phase-stable coaxial cables to minimise phase drift during field sweeps. Unless otherwise stated, measurements were performed with a microwave source power of \SI{0}{dBm} and an intermediate-frequency (IF) bandwidth of \SI{1}{kHz}. This power was chosen to remain in the linear-response regime and avoid nonlinear heating or multi-magnon scattering in the YIG sphere. Before measurement, a full two-port OSMT calibration was performed at the cable ends to remove electrical delay and insertion loss from the transmission lines. The static magnetic field $\mathbf H_{\mathrm{DC}}$ lies in the $yz$ plane at angle $\theta$ to the $z$ axis, as depicted in \figsub{fig:S1}{fig:S1a}.  We define $Z$ as the height of the sphere from the planar board; $Z=0$ places the sphere at the cavity midplane and increasing $Z$ raises it along $+\hat{\mathbf z}$. In this study, $Z$ is adjusted from \SI{0} to \SI{3}{mm} using a spacer stack comprising a \SI{1}{mm} PTFE block for coarse steps with \SI{0.05}{mm} Kapton layers for fine control. PTFE and Kapton are non-magnetic, low-loss dielectrics, so the spacer assembly does not introduce magnetic damping or materially reduce the cavity quality factor, while providing stable and repeatable height settings.

The resonator is fabricated on an RT/Duroid\textsuperscript{\textregistered}~6010LM substrate (thickness \SI{1.27}{mm}, relative permittivity $\varepsilon_r=10.2$), which permits a compact footprint at the operating frequency. A circular aperture at the cavity center accommodates the YIG sphere and registers it with the simulated maximum of the microwave magnetic field $|\mathbf H|$ for strongest magnon–photon coupling. Field confinement is provided by a via fence along the cavity perimeter, \figsub{fig:S1}{fig:S1b}, with center-to-center spacing \SI{1.15}{mm} (nine vias per side), forming the effective SIW sidewalls. The principal layout parameters are a square cavity of side $a=\SI{10.8}{mm}$, feedline width $w=\SI{1.75}{mm}$, via diameter $c=\SI{0.7}{mm}$, and central aperture $b=\SI{2}{mm}$.

%% file: sections/02-cst-fieldmaps/section.tex
\begin{figure}[h]
  \centering
  \includegraphics[width=0.6\linewidth]{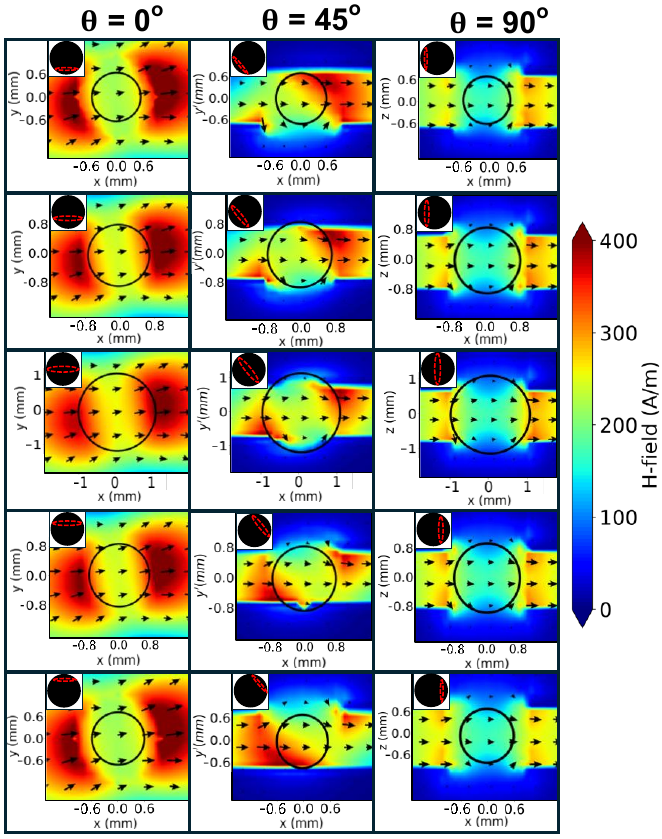}
  \caption[CST microwave magnetic field maps versus field angle and slice height.]
  {\justifying
  \textbf{CST microwave magnetic field maps.} Central cross-sections of a \SI{2}{mm}-diameter YIG sphere for
  $\theta=0^\circ,45^\circ,90^\circ$ (columns) and different slice heights through the sphere (rows).
  The color encodes $|\mathbf H|$; arrows show in-plane field direction and relative magnitude.
  The black circle marks the sphere boundary.
  For $\theta=0^\circ$ the section is the $xy$ plane; for $\theta=90^\circ$ it is the $xz$ plane; for $\theta=45^\circ$ it is the rotated plane $y'=(y+z)/\sqrt{2}$, giving a mixed projection of the cavity field.}
  \label{fig:S2}
\end{figure}

Figure~\ref{fig:S2} visualizes the distribution of the cavity's microwave magnetic field with $\theta$ within and around the YIG sphere. Each column corresponds to results for a different $\theta$ (0$^\circ$, 45$^\circ$ and 90$^\circ$), displaying the dynamic magnetic field strength for different cross-sections within the sphere as indicated by the red dotted curves in each inset. The 2D maps plot the field magnitude $|\mathbf H|$ with arrows indicating the local in-plane direction and relative strength. Important observation for these plots is that the strength of magnetic fields becomes weaker as you increase $\theta$ for the same cross section cuts, as one can compare three plots in the same row for different $\theta$. This explains the angular dependence of $\eta$ as we discuss in the main text.

%% file: sections/03-mode-id/section.tex
\noindent

Simulations of the hybrid magnon-photon coupling were performed using COMSOL Multiphysics\textsuperscript{®}, a finite element method (FEM) software solving coupled partial differential equations in arbitrary 2D/3D geometries.

The Radio Frequency (RF) module was used to model microwave cavities and waveguides, specifically the "Electromagnetic Waves, Frequency Domain" (\textit{emw}) interface, required for the frequency-dependent Polder tensor. This interface solves the time-harmonic wave equation $\nabla \times (\mu_r^{-1} \nabla \times \textbf{E}) - k_0^2 \epsilon_r \textbf{E} = 0$, with $\textbf{E}(\textbf{r}, t)= \textbf{E}(\textbf{r})e^{i \omega t}$, $\epsilon_r$ and $\mu_r$ the relative permittivity and permeability, and $k_0= \omega \sqrt{\epsilon_0 \mu_0}= \omega/c$. Two boundary conditions were used throughout: \textit{Scattering boundary conditions} ($\textbf{n}_2 \times (\textbf{E}_1 - \textbf{E}_2) = 0$, $\textbf{n}_2 \times (\textbf{H}_1 - \textbf{H}_2) = 0$) to simulate reflection-free open air domains, and perfect electric conductor (PEC) conditions ($\textbf{n} \times \textbf{E} = 0$) on cavity walls to reduce computational cost. Mesh element size followed the standard rule of 5-10 elements per wavelength. Mesh finesse was set locally depending on the physics.

Frequency Domain studies compute the complex \textit{S}-matrix using coaxial ports terminated with 50~$\Omega$ matched loads (Port~1 active/excitation, Port~2 passive/output), with
\begin{equation}
S_{11} = \frac{\int_{A_1}\left(\textbf{E} - \textbf{E}^{(1)}_j e^{i \phi}\right) \cdot \textbf{E}^{(1)^*}_j \, \text{d}A}{\int_{A_1} \left|\textbf{E}^{(1)}_j\right|^2 \text{d}A}, \qquad S_{21} = \frac{\int_{A_2} {\textbf{E}} \cdot \textbf{E}^{(2)^*}_j \, \text{d}A}{\int_{A_2} \left|\textbf{E}^{(2)}_j\right|^2 \text{d}A},
\end{equation}
where $\textbf{E}^{(n)}_j$ is the mode field $j$ on port $n$ of area $A_n$. Frequency Domain studies used the generalized minimal residual solver (GMRES) by default.

The spherical YIG sample was modeled via its demagnetization field $H_d=-M_s/3$, giving $H_{tot}=H_{\mathrm{DC}}-M_s/3$. As COMSOL Multiphysics\textsuperscript{®} does not support micromagnetics, the magnetic response was implemented as a complex Polder permeability tensor, derived from the linearized LLG equation under the rotating-wave and macrospin approximations:
\begin{equation}
\Bar{\bm{\mu}}=\mu_0(1+\Bar{\bm{\chi}}) = \mu_0
\begin{pmatrix}
1+\chi_{xx} & \chi_{xy} & 0 \\
-\chi_{yx} & 1+\chi_{yy} & 0\\
0 & 0 & 1
\end{pmatrix},
\end{equation}
with
\begin{align}
\chi_{xx} &= \frac{\omega_0\omega_M(\omega_0^2-\omega^2)+\omega_0\omega_M\omega^2\alpha^2}{(\omega_0^2-\omega^2(1+\alpha^2))^2+(2\omega_0\omega\alpha)^2} - i\frac{\alpha\omega\omega_M(\omega_0^2-\omega^2(1+\alpha^2))}{(\omega_0^2-\omega^2(1+\alpha^2))^2+(2\omega_0\omega\alpha)^2},\\
\chi_{xy} &= \frac{2\omega_0\omega_M\omega^2\alpha}{(\omega_0^2-\omega^2(1+\alpha^2))^2+(2\omega_0\omega\alpha)^2} + i\frac{\omega\omega_M(\omega_0^2-\omega^2(1+\alpha^2))}{(\omega_0^2-\omega^2(1+\alpha^2))^2+(2\omega_0\omega\alpha)^2},
\end{align}
$\chi_{yy} = \chi_{xx}$, $\chi_{yx} = \chi_{xy}$, $\omega_M=\gamma M_s$, $\omega_0=\gamma (H_{\mathrm{DC}}-M_s/3)$. Coupled simulations combined this tensor with a Frequency Domain study parametrically swept over $H_{\mathrm{DC}}$, using the parallel direct sparse solver in place of GMRES for improved robustness on strongly coupled multiphysics systems.

\begin{figure}[h]
  \centering
  \includegraphics[width=0.8\linewidth]{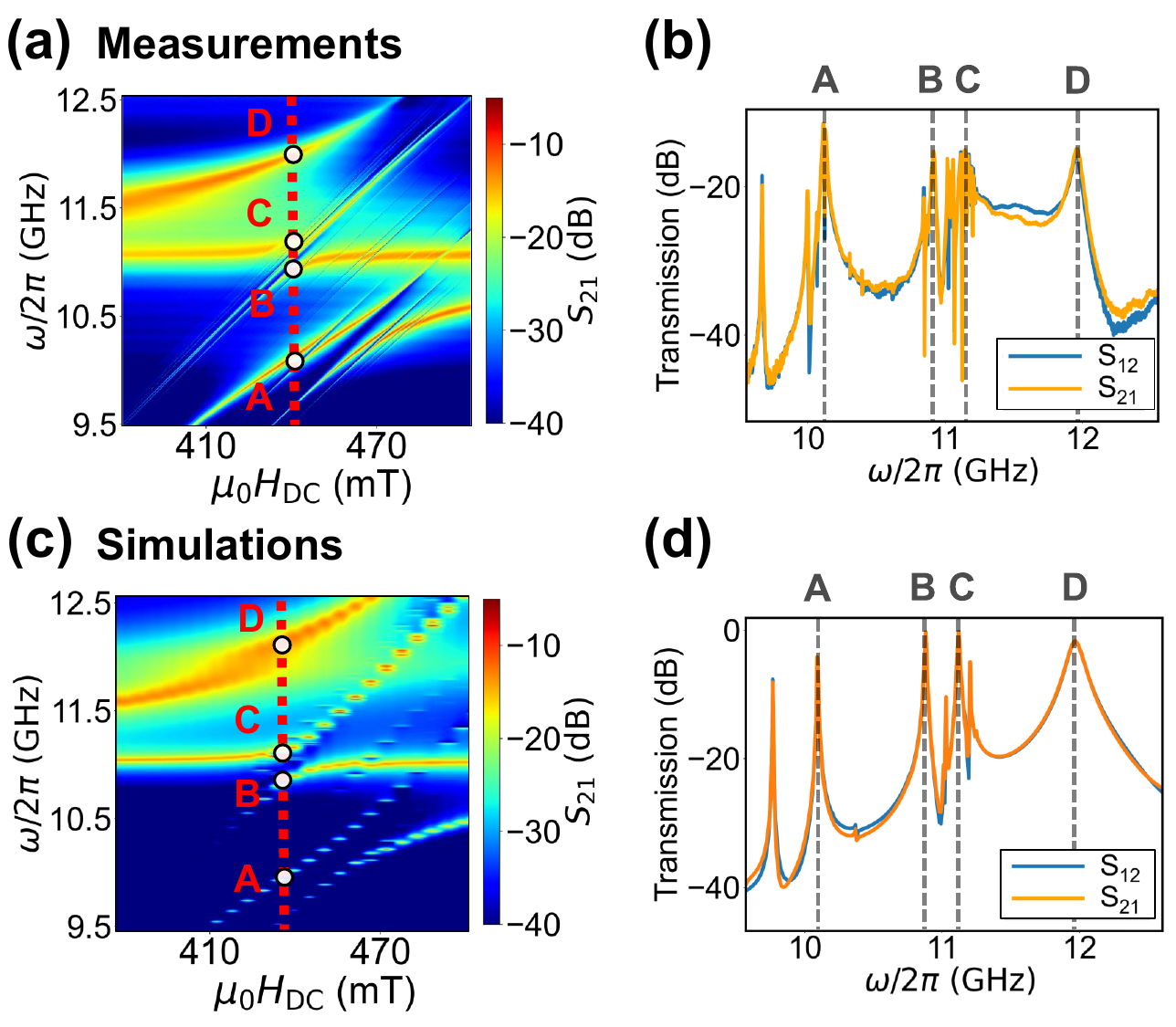}
  \caption[Measured and simulated double anticrossing and line cuts.]
  {\justifying
  \textbf{Double anticrossing and line cuts.}
  (a) Measured $S_{21}$ versus frequency and field, showing two anticrossings at the same bias field; branches are labeled A–D.
  (b) On-resonance line cuts of $S_{21}$ and $S_{12}$ at the dashed line in (a), revealing four peaks A–D.
  (c) Simulation for the same geometry and field range, reproducing the double anticrossing.
  (d) Simulated line cut at the on-resonance field (dashed line in (c)).}
  \label{fig:S3}
\end{figure}

\begin{figure}[h]
  \centering
  \includegraphics[width=0.8\linewidth]{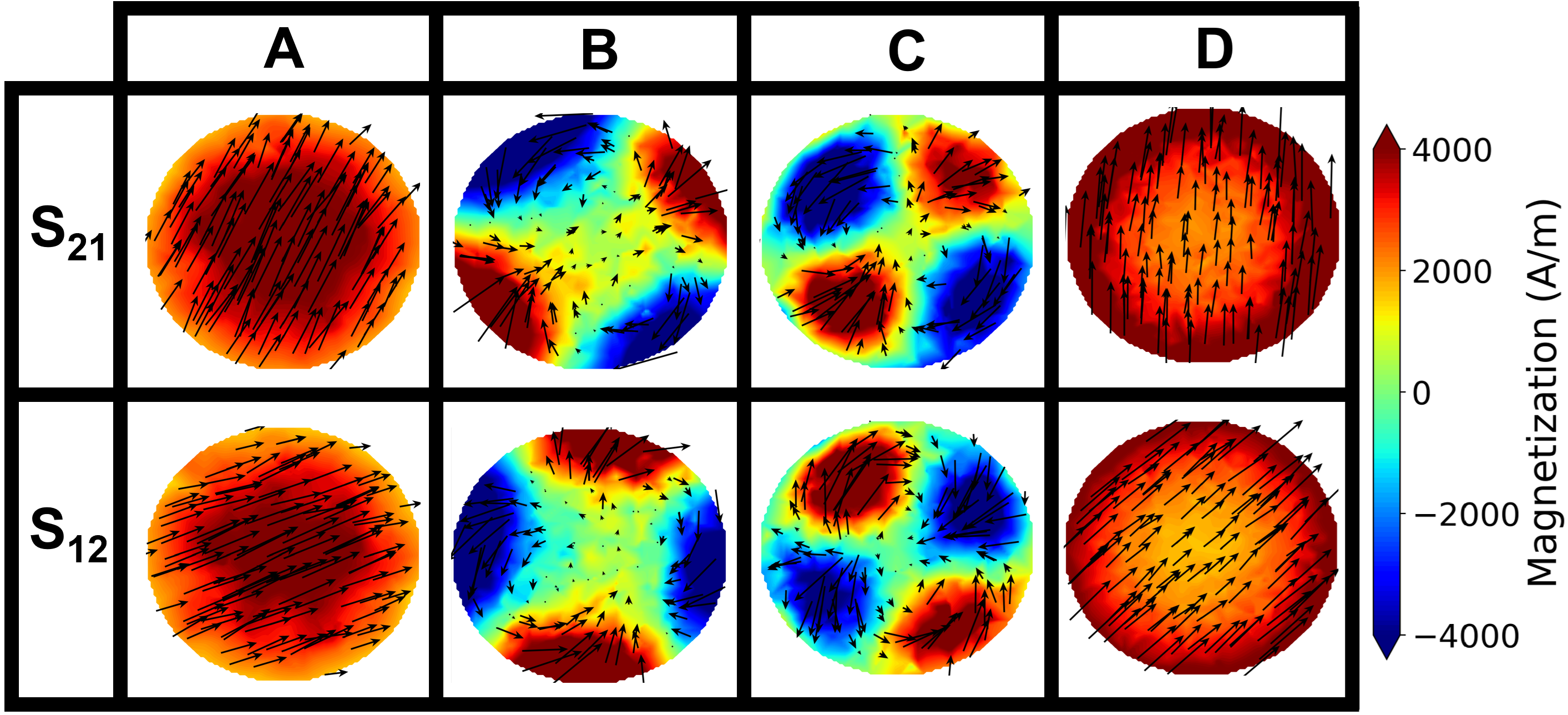}
  \caption[COMSOL dynamic magnetization patterns for branches A–D.]
  {\justifying
  \textbf{Dynamic magnetization for A–D.}
  COMSOL simulations of the central cross-section of the \SI{2}{mm} YIG sphere at the on-resonance field.
  Top row: excitation from port 2 ($S_{21}$). Bottom row: excitation from port 1 ($S_{12}$).
  The color encodes the magnetization component along the bias field; arrows indicate in-plane precession.}
  \label{fig:S4}
\end{figure}

\begin{figure}[h]
  \vspace{1.0em}
  \centering
  \includegraphics[width=0.85\linewidth]{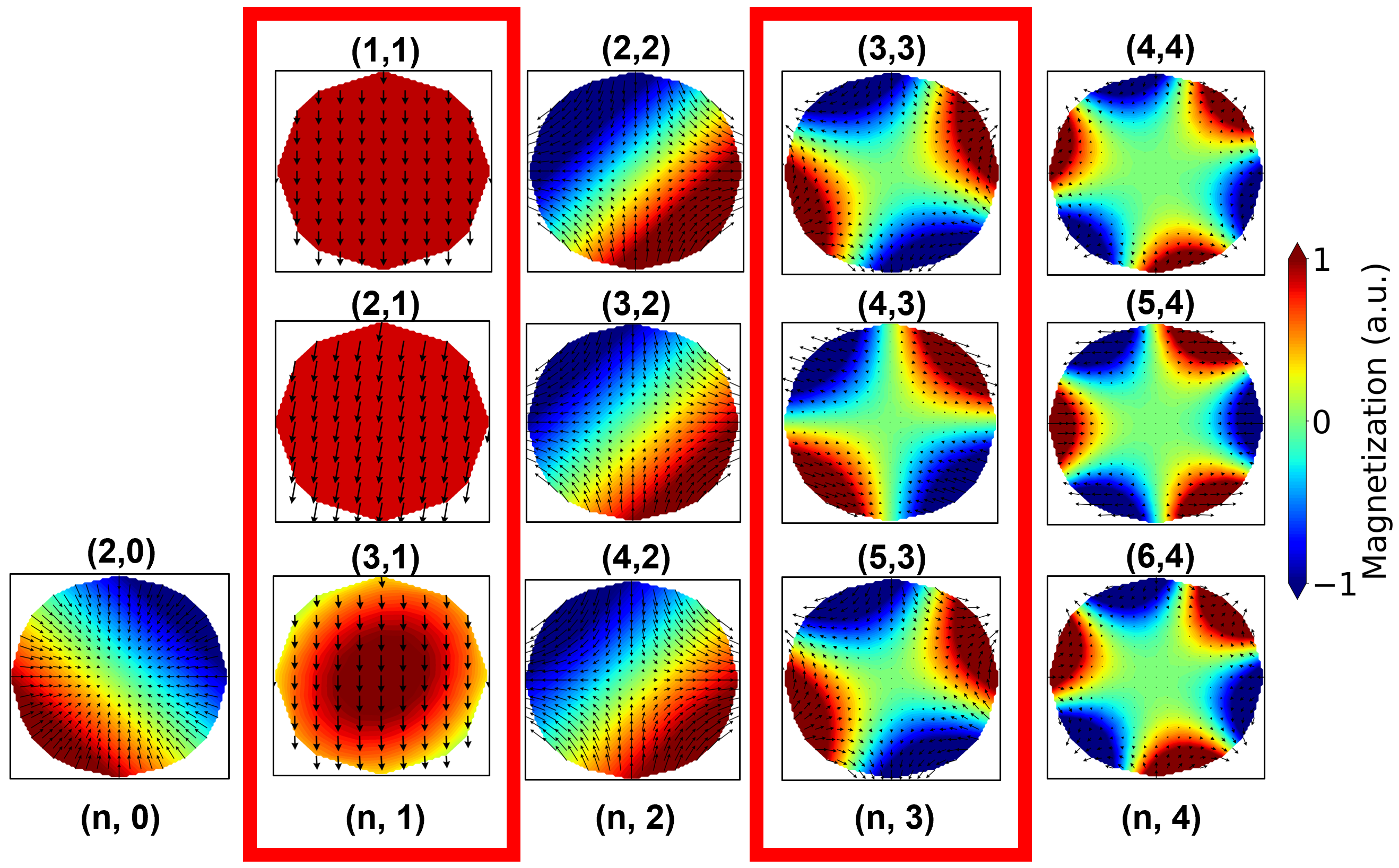}
  \caption[Analytical Walker-mode patterns used for mode assignment.]
  {\justifying
  \textbf{Analytical Walker-mode patterns.}
  Magnetization distributions for spherical Walker modes $(n,m)$ computed from the standard formulas of Ref.~\smcite{SM-FletcherBell1959}.
  The Kittel mode $(1,1)$ and representative higher-order $(n,3)$ modes are highlighted (red boxes) for comparison with the simulated profiles in Fig.~\ref{fig:S4}.}
  \label{fig:S5}
\end{figure}

\begin{figure}[h]
  \centering
  \includegraphics[width=0.6\linewidth]{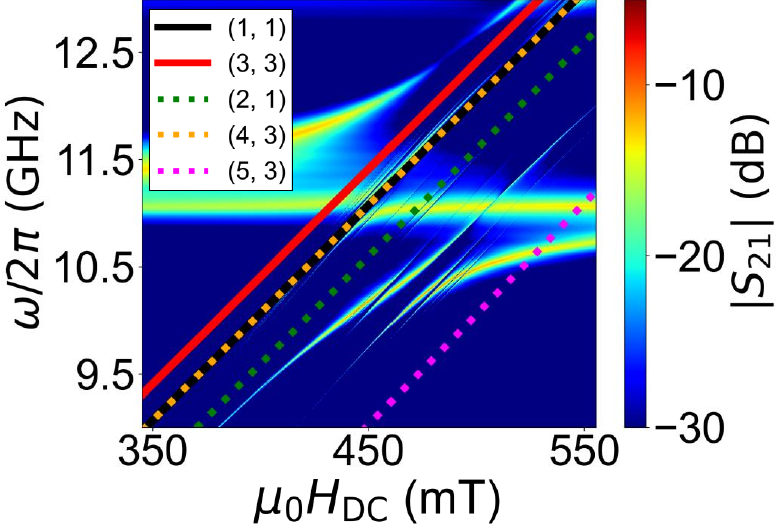}
  \caption[Measured spectrum overlaid with analytical Walker-mode dispersions.]
  {\justifying
  \textbf{Walker-mode dispersions overlaid on data.}
  Measured $|S_{21}|$ map with analytical resonance lines for selected mode families,
  calculated from the Walker-mode expressions of Ref.~\smcite{SM-FletcherBell1959}.
  Solid and dashed curves denote $(n,1)$ and $(n,3)$ families, respectively; specific modes are indicated in the legend.}
  \label{fig:S6}
\end{figure}

Figure~\ref{fig:S3} demonstrates that two distinct magnon resonances couple to the same chiral SIW cavity mode at a single bias field, producing the observed double anticrossing. The corresponding $S_{21}$/$S_{12}$ line cuts isolate four hybrid peaks (A–D), COMSOL simulations of the central cross-section of the 2 mm YIG sphere at the on-resonance field shown in Fig.~\ref{fig:S4} can help identify the magnon modes that couple to the cavity mode. The rotation sense follows that of cavity chirality, while the nodal structure distinguishes the magnon modes, i.e. (1,1) and (4,3), that couple under the two excitation directions. We then compared the spatial profiles of Fig.~\ref{fig:S4} with those obtained by analytical approaches shown in Fig.~\ref{fig:S5}, in order to understand their radial and azimuthal characters indexed by ($n,m$). Comparison of nodal topology and precession sense links the measured branches to the $(n,1)$ and $(n,3)$ families of Walker modes in a uniformly magnetized sphere~\smcite{SM-FletcherBell1959}. During this process, it is vital that the $(1,1)$ and $(4,3)$ modes are energetically degenerate as shown in Fig.~\ref{fig:S6}.


%% file: sections/04-gen-io/section.tex
\noindent
We model the coupled cavity--magnon system with coherent and dissipative
interactions using the von Neumann--Lindblad formalism
~\smcite{SM-Walls2008,SM-Metelmann_PRX2015},
\begin{equation}
\dot{\rho}
=
-\frac{i}{\hbar}[H_{\mathrm{coh}},\rho]
+\mathcal{L}\rho .
\label{eq:S_master}
\end{equation}

Within the rotating-wave approximation (RWA), the coherent Hamiltonian
is~\smcite{SM-Wang2019PRL}
\begin{equation}
\frac{H_{\mathrm{coh}}}{\hbar}
=
\omega_c a^\dagger a
+\omega_m b^\dagger b
+J(a^\dagger b+a b^\dagger),
\label{eq:S_Hcoh}
\end{equation}
where $a$ ($b$) denotes the annihilation operator of the cavity
(magnon) mode and $J$ is the coherent coupling strength.

The corresponding effective non-Hermitian Hamiltonian is
\begin{equation}
\frac{H_{\mathrm{eff}}}{\hbar}
=
\tilde{\omega}_c a^\dagger a
+\tilde{\omega}_m b^\dagger b
+\left(J-i\Gamma e^{i\Phi}\right)
(a^\dagger b+a b^\dagger),
\label{eq:S_Heff}
\end{equation}
where
$\tilde{\omega}_{c(m)}=\omega_{c(m)}-i\gamma_{c(m)}$
includes the intrinsic damping rates $\gamma_{c,m}$,
$\Gamma$ is the dissipative coupling strength, and $\Phi$ is the
direction-dependent phase associated with the shared radiative
channel. For the convention used here, the forward ($S_{21}$) and
backward ($S_{12}$) propagation directions are represented by
$\Phi=0$ and $\Phi=\pi$, respectively. In the unified model,
$\Gamma$ is not an independent coupling parameter but is determined
by the coupling of the cavity and magnon modes to the same radiative
bath.

Following the reservoir-engineering approach of
Ref.~\smcite{SM-Metelmann_PRX2015}, the coupling of the cavity and magnon
modes to the common travelling-wave reservoir is described by the
collective jump operator
\begin{equation}
\hat{o}
=
\sqrt{\kappa_c}\,\hat{a}
+
\sqrt{\kappa_m}\,e^{i\Phi}\hat{b},
\label{eq:S_jump}
\end{equation}
where $\kappa_c$ and $\kappa_m$ are the external coupling rates of
the cavity and magnon modes, respectively. Importantly, the same
direction-dependent phase $\Phi$ appears in both the common
radiative channel and the dissipative cavity--magnon interaction.

The total master equation is then
\begin{equation}
\dot{\rho}
=
-\frac{i}{\hbar}[H_{\mathrm{coh}},\rho]
+\mathcal{L}[\hat{o}]\rho
+\gamma_c\mathcal{L}[a]\rho
+\gamma_m\mathcal{L}[b]\rho,
\label{eq:S_master_total}
\end{equation}
where the collective dissipator describes coupling to the shared
radiative bath and the last two terms represent the intrinsic cavity
and magnon losses.

Expanding $\hat{o}^{\dagger}\hat{o}$ generates, in addition to the
direct radiative damping of the two modes, a cross term proportional
to $\sqrt{\kappa_c\kappa_m}\,e^{i\Phi}$. The dissipative coupling is
therefore constrained by the common-bath relation
\begin{equation}
\Gamma
=
R\sqrt{\kappa_c\kappa_m},
\label{eq:S_Gamma_commonbath}
\end{equation}

where $R$ is a real, direction-independent proportionality constant.
A joint fit across the $Z$-sweep, imposing a single common $R$ in
$\Gamma(Z)=R\sqrt{\kappa_c(Z)\kappa_m(Z)}$ for all sphere positions,
gives $R=2.98\pm0.13$, with the uncertainty obtained from propagation
of the fitted-parameter uncertainties.

With $\Gamma$ constrained by
Eq.~\eqref{eq:S_Gamma_commonbath}, the corresponding linearized
Langevin equations, neglecting quantum noise operators and considering
only the classical mean response relevant to microwave transmission
measurements, are
\begin{align}
\dot a
&=
\bigl[-i\omega_c-(\kappa_c+\gamma_c)\bigr]a
-
\bigl(iJ+e^{i\Phi}\Gamma\bigr)b
+\sqrt{\kappa_c}\,p_{\mathrm{in}},
\label{eq:S_LangA}
\\
\dot b
&=
\bigl[-i\omega_m-(\kappa_m+\gamma_m)\bigr]b
-
\bigl(iJ+e^{i\Phi}\Gamma\bigr)a
+\sqrt{\kappa_m}\,p_{\mathrm{in}},
\label{eq:S_LangB}
\end{align}
where $p_{\mathrm{in}}$ is the incident microwave field.

Fourier transforming
($\dot a\rightarrow-i\omega a$,
$\dot b\rightarrow-i\omega b$)
gives
\begin{align}
\bigl[i(\omega-\omega_c)-(\kappa_c+\gamma_c)\bigr]a
-
\bigl(iJ+e^{i\Phi}\Gamma\bigr)b
&=
-\sqrt{\kappa_c}\,p_{\mathrm{in}},
\label{eq:S_freqA}
\\
\bigl[i(\omega-\omega_m)-(\kappa_m+\gamma_m)\bigr]b
-
\bigl(iJ+e^{i\Phi}\Gamma\bigr)a
&=
-\sqrt{\kappa_m}\,p_{\mathrm{in}}.
\label{eq:S_freqB}
\end{align}

The input--output boundary condition follows from the same collective
jump operator in Eq.~\eqref{eq:S_jump}. The output field is therefore
written as
\begin{equation}
p_{\mathrm{out}}
=
p_{\mathrm{in}}
-i\sqrt{\kappa_c}\,a
-i\sqrt{\kappa_m}\,e^{i\Phi}b,
\label{eq:S_io_general}
\end{equation}
where $p_{\mathrm{out}}$ is the output microwave field amplitude.
The phase multiplying the magnon-channel contribution is therefore
the same $\Phi$ that appears in the dissipative interaction, and no
additional propagation or resonant phase parameters are introduced.

Solving Eqs.~\eqref{eq:S_freqA}--\eqref{eq:S_freqB} together with the
single-bath input--output relation in Eq.~\eqref{eq:S_io_general}, we
define
\begin{equation}
\Delta_c=\omega-\omega_c,
\qquad
\Delta_m=\omega-\omega_m,
\label{eq:S_detunings}
\end{equation}
and
\begin{equation}
X_c=\Delta_c+i\gamma_c,
\qquad
X_m=\Delta_m+i\gamma_m,
\qquad
W=J-i\Gamma e^{i\Phi}.
\label{eq:S_XW}
\end{equation}
The forward and backward transmission coefficients are then
\begin{equation}
S_{21(12)}
=
\frac{
W^2-X_cX_m
-\sqrt{\kappa_c\kappa_m}\left(1+e^{i\Phi}\right)W
+\kappa_c\kappa_m\left[1-i\left(1+e^{i\Phi}\right)\right]
-\kappa_c(1+i)X_m
-\kappa_m\left(e^{i\Phi}+i\right)X_c
}{
(\kappa_c+\gamma_c-i\Delta_c)
(\kappa_m+\gamma_m-i\Delta_m)
+\left(J-i\Gamma e^{i\Phi}\right)^2
}.
\label{eq:S_S21_general}
\end{equation}

Here $J$ denotes the coherent coupling strength, while the dissipative coupling $\Gamma$ is constrained by
Eq.~\eqref{eq:S_Gamma_commonbath}. The same direction-dependent phase $\Phi$ enters both the dissipative interaction and the magnon input--output channel, with $\Phi=0$ and $\Phi=\pi$ corresponding to the two propagation directions in the convention used here.

%% file: sections/05-angle-dep/section.tex
\begin{figure}[h]
  \centering
  \includegraphics[width=0.55\linewidth]{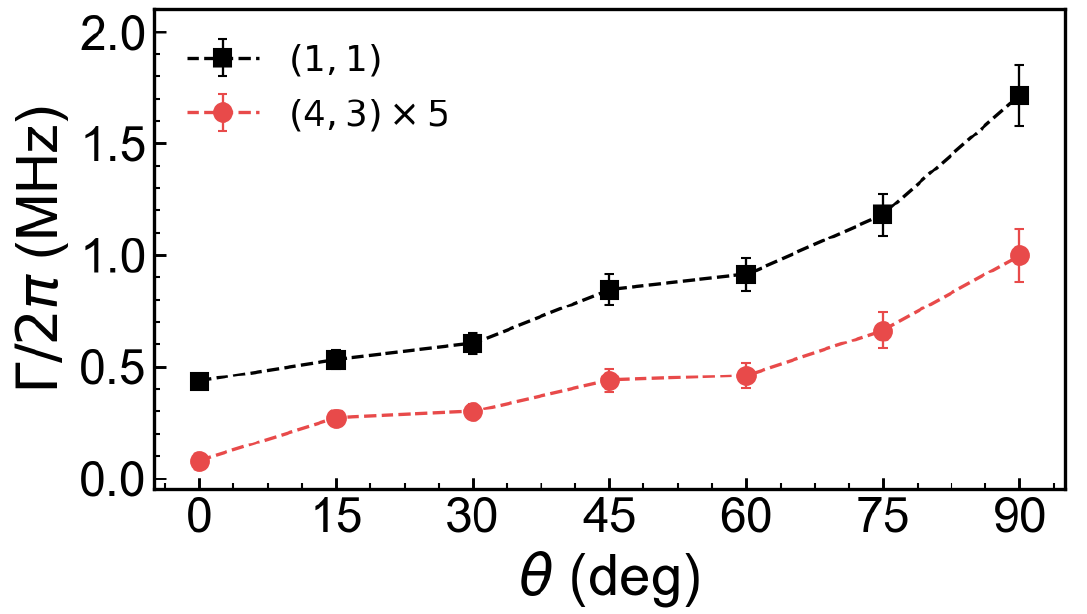}
  \caption[Angular dependence of the dissipative coupling.]
  {\justifying
  \textbf{Angular dependence of the dissipative coupling rate.}
  Dissipative coupling rate $\Gamma/2\pi$ versus the static magnetic
  field angle $\theta$ for the uniform $(1,1)$ mode (black squares)
  and the higher-order $(4,3)$ mode (red circles). The values for the
  $(4,3)$ mode are multiplied by a factor of 5 for visual clarity.
  Within the unified single-bath model, $\Gamma$ is evaluated from
  the common-bath relation
  (Eq.~S10).
  Error bars indicate the estimated uncertainties of the plotted
  values.}
  \label{fig:S7}
\end{figure}

\noindent
To verify that the angular evolution of the transmission spectra at
$Z=0$ is governed primarily by the coherent coupling $J$, we evaluate
the dissipative coupling rate $\Gamma/2\pi$ for both the $(1,1)$ and
$(4,3)$ magnon modes across the measured field angles $\theta$ using
the unified single-bath relation
introduced in
Eq.~\eqref{eq:S_Gamma_commonbath}. As shown in
Fig.~\ref{fig:S7}, $\Gamma/2\pi$ exhibits a moderate upward trend as
the magnetic field rotates toward the in-plane configuration
($\theta=90^\circ$). Nevertheless, its magnitude remains
considerably smaller than $J/2\pi$ throughout the measured angular
range. This confirms that the angular dependence of the hybridization at
$Z=0$ remains predominantly governed by the coherent
cavity--magnon coupling, with the dissipative contribution remaining
considerably smaller in magnitude throughout the angular range.

%% file: sections/06-height-dep/section.tex
\begin{figure}[!htb]
  \centering
  \includegraphics[width=\linewidth]{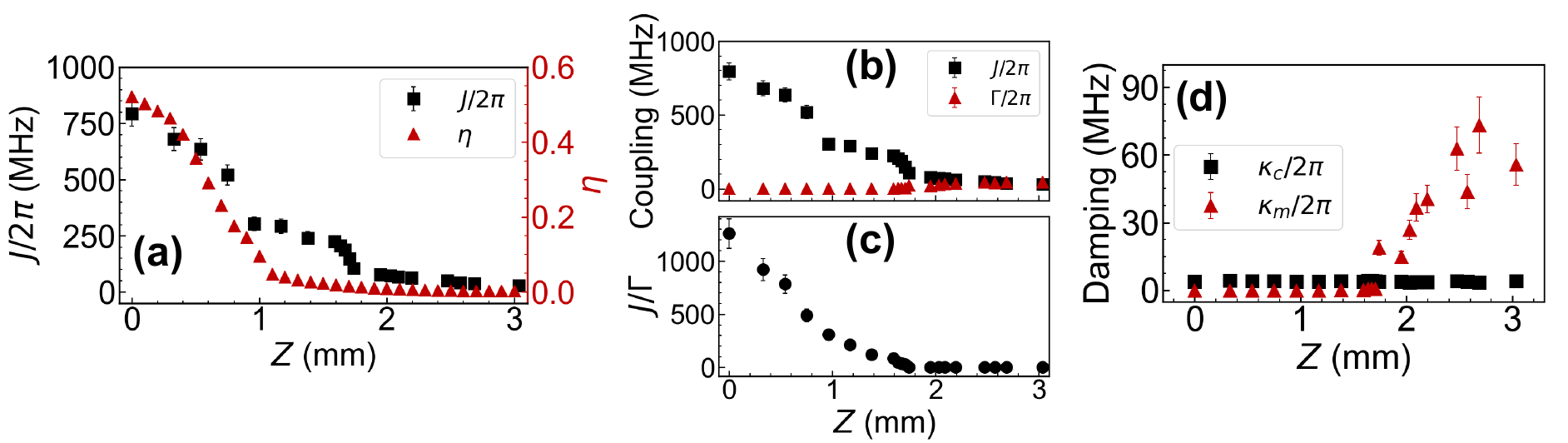}
  \setcounter{subfigure}{0}
  \subfiganchor{fig:S8a}
  \subfiganchor{fig:S8b}
  \subfiganchor{fig:S8c}
  \subfiganchor{fig:S8d}
\caption[Height dependence of the coupling parameters.]
{\justifying
\textbf{Height dependence of the coupling parameters.}
(a) Coherent coupling $J/2\pi$ (left axis) and calculated spatial
overlap factor $\eta$ (right axis) versus the YIG sphere vertical
displacement $Z$.
(b) Comparison of the coherent and dissipative coupling rates,
$J/2\pi$ and $\Gamma/2\pi$, versus $Z$.
(c) Coupling ratio $J/\Gamma$ versus $Z$.
(d) Cavity and magnon external coupling rates,
$\kappa_c/2\pi$ and $\kappa_m/2\pi$, versus $Z$.
Within the unified single-bath model, the dissipative coupling
$\Gamma$ is evaluated from the common-bath relation (Eq.~S10).
Error bars indicate the estimated uncertainties of the plotted
values.}
  \label{fig:S8}
\end{figure}

\noindent
To elucidate the physical mechanism behind the geometric tuning, we
characterize the height ($Z$) dependence of the system's coupling
parameters for the fundamental $(1,1)$ mode, as summarized in
Fig.~\ref{fig:S8}.

\figsub{fig:S8}{fig:S8a} shows that moving the YIG sphere away from
the on-chip cavity substantially reduces its overlap with the cavity
photon mode. Consequently, the spatial overlap factor $\eta$ decreases
rapidly with increasing $Z$, leading to a corresponding reduction in
the coherent coupling strength $J$. The evolution of the coherent and
dissipative coupling strengths is compared directly in
\figsub{fig:S8}{fig:S8b}. While $J$ decreases monotonically with
increasing $Z$, $\Gamma$ shows an overall increase. Within the unified
single-bath model, this dissipative coupling is constrained by
$\Gamma=R\sqrt{\kappa_c\kappa_m}$ rather than treated as an
independent coupling parameter. The contrasting evolution of $J$ and
$\Gamma$ therefore produces the strong reduction of $J/\Gamma$ with
increasing $Z$, as shown in \figsub{fig:S8}{fig:S8c}, bringing the
system toward the regime where the coherent and dissipative coupling
strengths become comparable.

The $Z$ dependence of the cavity and magnon external coupling rates,
$\kappa_c/2\pi$ and $\kappa_m/2\pi$, is shown in
\figsub{fig:S8}{fig:S8d}. While $\kappa_c/2\pi$ remains comparatively weakly dependent on $Z$, $\kappa_m/2\pi$ increases substantially as the sphere is displaced from the cavity center. Through the common-bath relation in Eq.~\eqref{eq:S_Gamma_commonbath}, this increase in
$\kappa_m$ accounts for the corresponding enhancement of the
dissipative coupling $\Gamma$ with increasing $Z$.